\documentclass[prl,twocolumn,tightenlines,preprintnumbers,floatfix,nofootinbib]{revtex4}
\usepackage{graphicx,amsfonts,amssymb,amsmath}

\def\be{\begin{equation}}
\def\ee{\end{equation}}
\def\ba{\begin{eqnarray}}
\def\ea{\end{eqnarray}}
\def\ge{\mathrel{\raise.3ex\hbox{$>$\kern-.75em\lower1ex\hbox{$\sim$}}}}
\def\la{\mathrel{\raise.3ex\hbox{$<$\kern-.75em\lower1ex\hbox{$\sim$}}}}

\def\simgt{\mathrel{\raise.3ex\hbox{$>$\kern-.75em\lower1ex\hbox{$\sim$}}}}
\def\simlt{\mathrel{\raise.3ex\hbox{$<$\kern-.75em\lower1ex\hbox{$\sim$}}}}

\newcommand{\bi}[1]{\bibitem{#1}}
\newcommand{\fr}[2]{\frac{#1}{#2}}

\newcommand{\nc}{\newcommand}

\nc{\gone}{\bar g_{\pi NN}^{(1)}}
\nc{\gzero}{\bar g_{\pi NN}^{(0)}}
\nc{\al}{\alpha}
\nc{\ga}{\gamma}
\nc{\de}{\delta}
\nc{\ep}{\epsilon}
\nc{\ze}{\zeta}
\nc{\et}{\eta}
\renewcommand{\th}{\theta}
\nc{\Th}{\Theta}
\nc{\ka}{\kappa}
\nc{\rh}{\rho}
\nc{\si}{\sigma}
\nc{\ta}{\tau}
\nc{\up}{\upsilon}
\nc{\ph}{\phi}
\nc{\ch}{\chi}
\nc{\ps}{\psi}
\nc{\om}{\omega}
\nc{\Ga}{\Gamma}
\nc{\De}{\Delta}
\nc{\La}{\Lambda}
\nc{\Si}{\Sigma}
\nc{\Up}{\Upsilon}
\nc{\Ph}{\Phi}
\nc{\Ps}{\Psi}
\nc{\Om}{\Omega}
\nc{\ptl}{\partial}
\nc{\del}{\nabla}
\nc{\ov}{\overline}
\nc{\newcaption}[1]{\centerline{\parbox{15cm}{\caption{#1}}}}

\def\beq{\begin{equation}}
\def\eeq{\end{equation}}
\def\bmat{\begin{displaymath}}
\def\emat{\end{displaymath}}
\def\bear{\begin{eqnarray}}
\def\eear{\end{eqnarray}}
\def\bery{\begin{array}}
\def\ery{\end{array}}
\def\bit{\begin{itemize}}
\def\eit{\end{itemize}}
\def\ben{\begin{enumerate}}
\def\een{\end{enumerate}}
\def\btab{\begin{tabular}}
\def\etab{\end{tabular}}
\def\btbl{\begin{table}}
\def\etbl{\end{table}}
\def\bfig{\begin{figure}[htb]}
\def\efig{\end{figure}}
\def\bpic{\begin{picture}}
\def\epic{\end{picture}}

\def\ga{\mathrel{\raise.3ex\hbox{$>$\kern-.75em\lower1ex\hbox{$\sim$}}}}
\def\la{\mathrel{\raise.3ex\hbox{$<$\kern-.75em\lower1ex\hbox{$\sim$}}}}
\def\gappeq{\mathrel{\rlap {\raise.5ex\hbox{$>$}}
{\lower.5ex\hbox{$\sim$}}}}
\def\lappeq{\mathrel{\rlap{\raise.5ex\hbox{$<$}}
{\lower.5ex\hbox{$\sim$}}}}

\def\gyr{{\rm \, G\kern-0.125em yr}}
\def\mev{{\rm \, Me\kern-0.125em V}}
\def\gev{{\rm \, Ge\kern-0.125em V}}
\def\tev{{\rm \, Te\kern-0.125em V}}

\begin{document}

\preprint{\hspace*{1.5in}hep-ph/0510254$\;\;\;\;\;\;\;$CERN-PH-TH-2005-203\hspace*{2.23in}}

\setcounter{page}{1}

%\vspace*{0.1in}

\title{Flavor and \boldmath{$CP$} violating physics from new supersymmetric thresholds}

\author{Maxim Pospelov$^{\,(a,b)}$, 
Adam Ritz$^{\,(b,c)}$ and Yudi Santoso$^{\,(a,b)}$}

\affiliation{$^{\,(a)}${\it Perimeter Institute for Theoretical Physics, Waterloo,
Ontario N2J 2W9, Canada}\\
$^{\,(b)}${\it Department of Physics and Astronomy, University of Victoria, 
     Victoria, BC, V8P 1A1 Canada}\\
$^{\, (c)}${\it Theoretical Division, Department of Physics, CERN,
Geneva 23, CH-1211 Switzerland}}

\begin{abstract}

Treating the MSSM as an effective theory, we study the implications of having dimension 
five operators in the superpotential for flavor and $CP$-violating processes, exploiting 
the linear decoupling of observable effects with respect to the new threshold scale $\Lambda$. 
We show that the assumption of weak scale supersymmetry, when combined 
with the stringent limits on electric dipole moments and lepton flavor-violating processes, provides 
sensitivity to $\Lambda$ as high as $10^7-10^9$ GeV, while the next generation of experiments 
could directly probe the high-energy scales suggested by neutrino physics. 

\end{abstract}

\maketitle

\newpage

Weak-scale supersymmetry (SUSY) is a theoretical framework that helps to soften the 
so-called gauge hierarchy problem by removing the power-like ultraviolet sensitivity of 
the dimensionful parameters in the Higgs potential. It also has other advantages, notably
an improvement in gauge coupling unification and a natural dark matter candidate, which have
made it the standard paradigm for physics beyond the Standard Model (SM). However, the simplest
scenario -- the minimal supersymmetric standard model (MSSM) -- suffers from a number of well-known
tuning problems, due in part to the large array of possible parameters responsible for soft SUSY breaking \cite{susy}, and 
consequently the possibility of catastrophically large flavor and $CP$ violating amplitudes.
The absence of new flavor structures and order-one sources of $CP$-violation in the soft breaking sector, as evidenced respectively
by the perfect accord of the observed $K$ and $B$ meson mixing and decay with the predictions of the SM \cite{ut}
and the null results of electric dipole moment (EDM) searches \cite{Tl,Hg,n}, 
motivates continuing work on the specifics of SUSY breaking.

In the present Letter we will instead ask, given a solution to the flavor and $CP$ problems in the soft-breaking sector,
what sensitivity do we have to new high-scale sources of flavor and $CP$-violation? Such effects would arise 
through SUSY-preserving higher-dimensional operators generated at a new threshold $\La \gg M_W$. 
Such thresholds
are indeed expected due to various completions of the MSSM, e.g. via mechanisms for SUSY breaking and mediation, 
the breaking of flavor symmetries, and moreover via the physics generating neutrino masses and mixings. 
Intermediate scales are also suggested by the axion solution to the strong $CP$ problem, SUSY leptogenesis 
scenarios, and more entertainingly as a lowered GUT/string scale arising from large compactification radii of 
extra dimensions. % \cite{LED}. 
In contrast to nonuniversal or complex soft-breaking terms, 
the flavor and $CP$-violating observables induced by such operators will scale 
as $(\Lambda m_{\rm susy})^{-1}$, and thus the constraints on nonminimal flavor or $CP$ translate directly
into sensitivity to $\La$ far above the scale of the superpartner masses, $m_{\rm susy}$. 

At dimension five there are several well-known $R$-parity conserving operators associated with neutrino masses, 
$H_u L H_u L$, and baryon number violation, $U U D E$, $QQQL$ \cite{WeinbergSY}. 
The constraints on proton decay put severe restrictions on the size of 
baryon-number violating operators, $\Lambda_{b}  > 10^{24}$~{\rm GeV}, where $1/\Lambda_b$ 
is the overall normalization scale for these operators.  The ``super-seesaw''
operator $H_u L H_u L$
is a welcome addition to the MSSM superpotential, as it generates Majorana masses 
and mixing for neutrinos,
% \cite{neutrinos}, 
which imply $\Lambda_{\nu} \sim (10^{14} - 10^{16})~{\rm GeV}$.  Note that in the seesaw scenario, the actual scale
of right-haned neutrinos, $M_R$, is lower than $\La_\nu$, since $\Lambda_\nu^{-1} = Y_\nu^2M_R^{-1}$ with a small $Y_\nu$, 
as is also favored by SUSY leptogenesis.% \cite{leptogen}.

In what follows, we analyze in detail the remaining 
operators allowed in the $R$-parity conserving MSSM at dimension five level \cite{WeinbergSY}.
We write the superpotential as
\bear
{\cal W} &=& {\cal W}_{\rm MSSM} + \fr{y_h}{\Lambda_{h}}H_dH_uH_dH_u +
\fr{Y^{qe}_{ijkl}}{\Lambda_{qe}}(U_i Q_j )E_k L_l \nonumber\\
 && \!\!\!\!\!\!\!\!\!\!\!\!\! +\fr{Y^{qq}_{ijkl}}{\Lambda_{qq}}(U_iQ_{j}) (D_k Q_{l} )+
\fr{\tilde Y^{qq}_{ijkl}}{\Lambda_{qq}}(U_it^AQ_{j}) (D_kt^AQ_{l}),\label{qule}
\eear
where $y_h$, $Y_{qe}$, $Y_{qq}$ and $\tilde Y_{qq}$ are dimensionless coefficients,
the latter three being tensors in flavor space. 
The parentheses in (\ref{qule}) denote a contraction of colour indices. 
Note that since we will only consider supersymmetric thresholds, the superfield equations of
motion can be used to eliminate all dimension five corrections to the K\"ahler potential, e.g.
$K^{(5)} = c_{u}QUH_d^\dagger$, absorbing them in ${\cal W}^{(5)}$ and the Yukawa terms, and 
slightly modifying the soft-breaking sector. A renormalizable realization of 
(\ref{qule}) can easily be obtained, {\em e.g.} the MSSM extended by a singlet $N$ (the NMSSM) or an 
extra pair of heavy Higgses.
%(soft SUSY breaking would correct this picture only slightly \cite{H*}).

The full Lagrangian descending from (\ref{qule}) is rather cumbersome, and 
we will focus our attention here on those dimension five operators  
which are of potential phenomenological interest, specifically those that involve two SM fermions and 
two sfermions. We then proceed to integrate out the sfermions to obtain operators composed from the SM
fields (or more precisely those of a type II two-Higgs doublet model). We will impose the requirements of
flavor triviality and $CP$ conservation in the soft-breaking sector. Thus all dimension $\leq 4$ 
coefficients in the Higgs potential, trilinear terms $A_i$, gaugino masses $M_i$, and the $\mu$-parameter,
will be taken real. We will also make the simplifying assumption of universal sfermion masses, 
denoted $m_{\rm sq}$, $m_{\rm sl}$, which we will take, along
with $\mu$, $M_i$, to be somewhat larger than $M_W$. Deferring the full details \cite{prs2},
we quote the relevant results below:

{\em Correction to the SM fermion masses:}
The SM operators of lowest dimension that are of phenomenological interest are the 
fermion mass operators. From the diagrams of Fig.~1a, we obtain the following corrections:
\begin{eqnarray}
\label{delta_m}
\de (M_e)_{ij} &=& Y^{qe}_{klij}(M_u^{(0)})^*_{kl}
~\fr{3\ln(\Lambda_{qe}/m_{\rm sq})}{8\pi^2 \Lambda_{qe}}
(A_u^* + \mu \cot\beta) \nonumber\\
\de (M_d)_{ij} &=& K^{qq}_{klij}(M_u^{(0)})^*_{kl}
~\fr{\ln(\Lambda_{qq}/m_{\rm sq})}{4\pi^2 \Lambda_{qq}}
(A_u^* + \mu \cot\beta),\;\;\;\;\;
\end{eqnarray} 
with a similar correction to $M_u$. The notation implies summation over the repeated flavor indices, and 
we have defined the combination $K^{qq}\equiv (Y^{qq}-2\tilde{Y}^{qq}/3)$. $M^{(0)}_{e,d,u}$ denote unperturbed 
mass matrices arising from dimension four terms in the superpotential. 
Note that the corrections proportional to $A_{u}$ directly break SUSY, while those proportional to $\mu$ arise
from corrections to the K\"ahler potential. 

\begin{figure}
\centerline{\includegraphics[width=8.2cm]{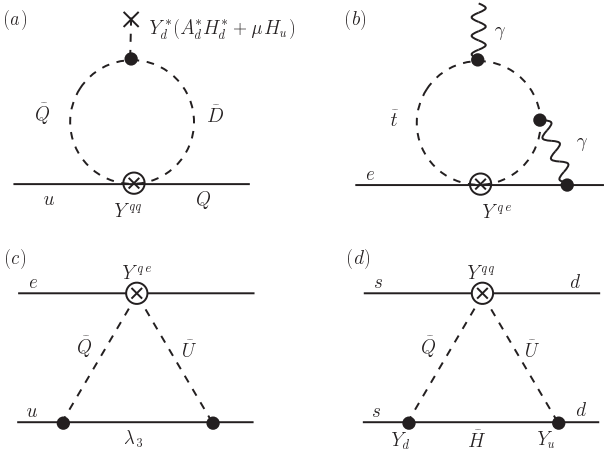}}
 \caption{\footnotesize Several representative loop corrections to: (a) SM fermion masses; 
(b) dipole amplitudes contributing to EDMs (cf. the supersymmetric Barr-Zee diagrams \cite{CKP}), 
$\mu\to e\gamma$, $b\to s\gamma$, $(g-2)_\mu$; and (c,d) dimension six four-fermion operators. 
The crossed vertex descends from dimension five terms in the superpotential (\ref{qule}).}
\label{f2} 
\end{figure}

{\em Dipole operators:}
At dimension five, dipole operators first arise at two-loop order, as in Fig.~1b.
In the charged lepton sector they result in 
\be
\label{dipole}
{\cal L}_{e} =    
\fr{A_u +\mu\cot\beta}{\Lambda^{qe}m_{\rm sq}^2} \fr{e\alpha}{12\pi^3} %\\\nonumber
(M_u)^*_{kl}Y^{qe}_{klij}\bar E_i (F\sigma) P_L E_j +(h.c.),
\ee
where we treated $LR$ squark mixing as a mass insertion, and used $P_L = \fr{1-\gamma_5}{2}$ and 
$(F\sigma) = F_{\mu\nu}\sigma^{\mu\nu}$. 
In the quark sector the corresponding results are more cumbersome due to 
a large number of possible diagrams.

Jumping an additional dimension, we now consider dimension six four-fermion 
operators generated by various terms in (\ref{qule}). Two representative
diagrams are shown in Fig.~1c,d.

{\em Semileptonic operators:}
Integrating out gauginos and sfermions as in Fig.~1c, we find the 
following semileptonic operators, sourced by $QULE$,
\be
{\cal L}_{qe}=
\fr{1}{\Lambda_{qe}m_{\rm susy}}\fr{\alpha_s}{3\pi} 
 Y^{qe}_{ijkl}\bar U_i Q_j \bar E_k L_l   + (h.c.).
\label{qqll}
\ee
Here $m_{\rm susy}^{-1}$ denotes a combination of 
superpartner masses folded with a loop function $F$: 
$m_{\rm susy}^{-1} = M_3m_{\rm sq}^{-2}F(M_3^2/ m_{\rm sl}^2)$, and 
$F(a) = 2~\fr{1-a +a\ln(a)}{(1-a)^2}$ with $F(1)=1$ (see \cite{masscorr} for the unequal mass 
case).
In (\ref{qqll}) we have retained only  the gluino-squark contribution,
which is expected to dominate unless there are additional hierarchies between 
the masses of sleptons and squarks.

{\em Four-quark operators:}
Integrating out gluinos and squarks as in Fig.~1c, we arrive at the following four-quark 
effective operators:
\begin{eqnarray}
{\cal L}_{qq} & = & \fr{1}{\Lambda_{qq}m_{\rm susy}}\fr{\alpha_s}{12\pi}
\;\;\;\;\;\;\;\;\;\;\;\;\;\;\;\; \label{qqqq}\\ 
&& \!\!\!\!\!\!\!\!\!\!\!\!\!\!\!\!\!\!\!\! \times 
K^{qq} \left[\frac{8}{3}(\bar U Q) (\bar D Q)+ (\bar U t^A Q) (\bar D t^A Q)\right]
 + (h.c.),\nonumber
\end{eqnarray}
where  the summation over flavor is carried out exactly as in (\ref{qule}). The largest
down-type $\De F=2$ operator arises instead from Fig.~1d, 
\begin{eqnarray}\label{dddd}
{\cal L}_{dd} &=& 
\fr{1}{\Lambda_{qq}m_{\rm susy}}\fr{1}{16\pi^2}
(Y^*_u)_{im}(Y^*_d)_{nj} K^{qq}_{ijkl} \\ 
 && \nonumber \!\!\!\!\!\!\!\!\!\!\!\!\!\!\!\!\!\!\!\! 
\times\left[\fr{1}{3} (\bar Q_m D_n) (\bar D_k Q_l)-(\bar Q_m t^A D_n) (\bar D_k t^A Q_l)\right]
  + (h.c.),
\end{eqnarray}
which inevitably contains additional Yukawa suppression originating from the 
Higgsino-fermion-sfermion vertices. Here $m_{\rm susy}$ is a combination of SUSY masses as in 
(\ref{qqll}) and (\ref{qqqq}) with $M_3$ replaced by $\mu$.

We will now turn to the phenomenological consequences and the sensitivity to $\Lambda^{qe}$ 
and $\Lambda^{qq}$ in various experimental channels. Of course, one of the most important 
issues is the flavour structure of the new couplings constants, $Y^{qe}$, $Y^{qq}$ and $\tilde Y^{qq}$. 
We will assume that these coefficients are of order one, and {\em do not factorize}: $Y^{qe} \neq Y_u Y_e$. 
With this assumption, we should first determine the natural scale for $\La$ such that the corrections to 
SM fermion masses do not exceed their measured values. 

{\em Particle masses and $\theta$-term:}
Taking $(M_u A_u)_{kl} =  (M_u A_u)_{33} \sim m_tA_t \sim 175 {\rm GeV} \times 300 {\rm GeV}$
%and using the expression for $\Delta m_e$ 
in (\ref{delta_m}), and assuming a maximal $Y^{qe}_{3311}\sim O(1)$, we arrive at the %following 
estimate,
\be
\Delta m_e \sim \fr{3 m_t A_t Y^{qe}_{3311}\ln(\Lambda^{qe}/m_{\rm sq})}{8\pi^2\Lambda^{qe}}
\sim 1 {\rm MeV} %\times ~
\fr{10^7{\rm GeV}}{\Lambda^{qe}}.
\label{dme}
\ee
Eq.~(\ref{dme}) clearly implies that the natural scale for new physics 
encoded in the semileptonic %dimension five 
operators in the superpotential 
is $\Lambda^{qe}\sim 10^7$ GeV, while the corresponding scale in the 
quark sector is slightly lower. % $\Lambda^{qq}\sim 10^6$ GeV. 

A strikingly high naturalness scale emerges from
consideration of the effective shift of $\bar \th$ due to the mass corrections (\ref{delta_m}). 
Assuming uncorrelated phases between $Y^{qq}$ and the eigenvalues of $Y_u$ and $Y_d$,
we find,
\be
\Delta \bar\theta \sim \fr{{\rm Im~}m_d}{m_d} \sim 
\fr{ {\rm Im~}K^{qq}_{3311}m_tA_t\ln(\Lambda^{qq}/m_{\rm sq})}{4\pi^2 m_d\Lambda^{qq}}
 \sim 
\fr{10^{7}~{\rm GeV}}{\Lambda^{qq}}.
\label{delta_th}
\ee
Eq.~(\ref{delta_th}) translates directly to an extremely strong bound on $\Lambda^{qq}$ in scenarios 
where $\bar \theta\simeq 0$ is engineered by hand, either by using discrete symmetries at high energies 
\cite{discrete} or by imposing an approximate global $U(1)$ symmetry at tree level to ensure $m_u^{(0)}=0$.  
In these cases, the experimental bound on the neutron EDM, $|d_n|  <  6\times 10^{-26} e\, {\rm cm}$ \cite{n} 
(soon to be updated \cite{n2}), combined with standard estimates for $d_n(\bar\theta)$ \cite{PR2005}
implies remarkable sensitivity to scales $\Lambda^{qq} \sim 10^{17}$ GeV. Future progress in EDM searches
(both for neutrons and heavy atoms) can bring this up to the Planck scale and beyond. In contrast, no constraints 
from (\ref{delta_th}) ensue within the axion scenario.

{\em Electric dipole moments from four-fermion operators:}
Electric dipole moments (EDMs) of neutrons and heavy atoms and molecules are the
primary probes for sources of flavor-neutral $CP$ violation \cite{PR2005}. 
In addition to $d_n$, the strongest constraints on $CP$-violating parameters arise 
from the atomic EDMs of thallium, $|d_{\rm Tl}| < 9 \times 10^{-25} e\, {\rm cm}$ \cite{Tl}, 
and mercury, $|d_{\rm Hg}| < 2   \times 10^{-28}  e\, {\rm cm}$ \cite{Hg}.

Assuming that $\bar\theta$ is removed by an appropriate symmetry, EDMs are mediated by higher-dimenional operators and
both (\ref{qqll}) and (\ref{qqqq}) are capable of inducing atomic/nuclear EDMs if the overall coefficients contain an extra 
phase relative to the quark masses. Restricting Eq.~(\ref{qqll}) to the first generation, we find the 
following $CP$-odd operators (with real $m_e$, $m_u$):
\be
{\cal L}_{CP} = -
\fr{\alpha_s{\rm Im}Y^{qe}_{1111}}{6\pi\Lambda_{qe}m_{\rm susy}}
\left[(\bar uu )\bar ei\gamma_5 e + (\bar ui\gamma_5u) \bar e e \right].
\ee
Accounting for QCD running from the SUSY scale to $1$GeV, 
and using the hadronic matrix elements over nucleon states,
$\langle N| (\bar uu + \bar dd)/2|N\rangle \simeq 4 \bar{N}N$ and 
$\langle n|\bar ui \gamma_5u |n\rangle \simeq -0.4(m_N/m_u)\bar n i \gamma_5 n$, 
we determine the induced corrections to the $CP$-odd electron-nucleon Lagrangian, 
${\cal L} = C_S \bar NN \bar e i \gamma_5 e + C_P \bar Ni \gamma_5N \bar e  e $,
\be
 C_S \sim \frac{2\times 10^{-4}}{1{\rm GeV}\times\Lambda^{qe}}, \;\;\;
 C_P \sim \frac{4\times 10^{-3}}{1{\rm GeV}\times\Lambda^{qe}}, \label{CSCP}
\ee
using maximal Im$Y^{qe}$ and taking $m_{\rm susy} = 300~{\rm GeV}$.

Comparing (\ref{CSCP}) to the limits on $C_S$ and $C_P$ deduced from 
the Tl and Hg EDM bounds \cite{PR2005}, we obtain the following sensitivity,
\begin{eqnarray}
\label{cslimit}
\Lambda^{qe} &\ga& 3 \times 10^8 ~{\rm GeV} ~~~~~~~~~~{\rm from~ Tl~ EDM} \\
\Lambda^{qe} &\ga& 1.5 \times 10^8 ~{\rm GeV} ~~~~~~~~\!{\rm from~ Hg~ EDM}\\
\Lambda^{qq} &\ga& 3\times 10^7 ~{\rm GeV} ~~~~~~~~~~{\rm from~ Hg~ EDM}.
\end{eqnarray}
The last relation results from sensitivity to the $CP$ violating operators $(\bar d i \gamma_5 d)(\bar u u)$
from (\ref{qqqq}), leading to the Schiff nuclear moment and the Hg EDM. 
These are remarkably large scales, and indeed not far below the scales suggested by neutrino physics.
In fact, the next generation of atomic/molecular EDM experiments \cite{nextedm} may reach sensitivities
sufficient to push $\La^{qe}$ into regions close to the suggested scale of  right-handed neutrinos.

Semileptonic operators involving heavy quark superfields are in turn strongly constrained
via two-loop corrections (\ref{dipole}) to the dipole amplitudes. The bound on $d_{\rm Tl}$
implies $|d_e|\la 1.6\times 10^{-27} e\, {\rm cm}$, which for maximal ${\rm Im}Y^{qe}_{1133}$ implies:
\be
\Lambda^{qe}\ga 1.3 \times 10^8{\rm GeV}.
\ee
Results analogous to (\ref{dipole}) apply for the quark EDMs and color EDMs,
furnishing a similar sensitivity to $\Lambda^{qq}$.

{\em Lepton flavour violation:}
Searches for lepton-flavour violation (LFV), such as $\mu\to e\gamma$ decay, 
and $\mu\to e$ conversion in nuclei, have resulted in stringent upper bounds on the 
corresponding branching ratio, ${\rm Br}({\mu\to e\gamma}) < 1.2\times 10^{-11}$ \cite{muegamma}, and the rate 
of conversion normalized on capture rate, $R({\mu \to e^-~  {\rm on ~Ti}}) < 4.3\times 10^{-12}$ \cite{sindrum},
with further improvement anticipated.
The latter bound implies a particularly high sensitivity to the semileptonic operators in (\ref{qule}).
The conversion is mediated by $(\bar uu)\bar e i \gamma_5 \mu$
and $(\bar uu) \bar e \mu$, and involves the same matrix elements 
as $C_S$. Using bounds on such scalar operators
derived elsewhere (see {\em e.g.} \cite{faessler}),
we conclude that $\mu\to e$ conversion probes energy scales as high as 
\be
\Lambda^{qe} \ga 1\times 10^8 {\rm GeV}~~~~~{\rm from}~{\mu^- \to e^-~  {\rm on ~Ti}}.
\label{muelimit}
\ee
The constraint on $\mu\to e\gamma$ probes similar, but slightly lower, scales as it requires 
a two-loop diagram as in Fig.~1b.
Disregarding an ${\cal O}(1)$ factor between (\ref{cslimit}) and (\ref{muelimit}), we conclude that 
searches for EDMs and LFV probe these extensions of the MSSM up to comparable energy scales 
of $\sim10^8$ GeV.

%{\em $K$ and $B$ meson mass difference:}
{\em Hadronic flavor constraints:}
Often, the most constraining piece of experimental information comes from the 
contribution of new physics to the mixing of neutral mesons, $K$ and $B$. However, in the 
present case, there is necessarily a significant loop and Yukawa suppression arising from (\ref{dddd}),
and the sensitivity is correspondingly weakened.
Taking $(\De m_K)_{\rm exp} \simeq 3.5\times 10^{-6} {\rm eV}$ \cite{pdg}, we 
find $\La^{qq} \ga (\tan\beta/50)\times 200\, {\rm GeV}$ 
\cite{prs2}. $\Delta m_B$ exhibits a similar sensitivity, while $\ep_K$ is about three orders of magnitude more
sensitive, but still well below the scales probed by EDMs and LFV. In contrast, it is clear that
these observables provide much better sensitivity to SUSY dimension-six operators, which impose no
additional suppression factors. Denoting the corresponding scale as $\La'$, we find 
%$\Delta m_K \sim \fr{0.25 {\rm ~ GeV}^3}{\mu'^2} \qquad \Longrightarrow \qquad \mu' \ga 8\times 10^6~{\rm GeV}$,
$\La' \ga 8\times 10^6~{\rm GeV}$, while $\epsilon_K$ is sensitive to scales 
$\sim  10^8$ GeV. 

Two-loop contributions to $b\rightarrow s\gamma$ (as in Fig.~1b) are not Yukawa suppressed and, with the current precision 
$\De Br(B\rightarrow X_s\gamma) \sim 10^{-4}$ \cite{pdg}, are somewhat more sensitive. We find
$\La^{qq} \ga 10^3-10^4 {\rm GeV}$ (for $Y^{qq}_{3233}\sim 1$), still well below the 
sensitivity in other channels.

{\em Constraints on the Higgs operator:}
The high sensitivity to $QULE$ and $QUQD$ arises primarily because they can flip the light fermion chirality 
without Yukawa suppression. It would then come as no surprise if $H_uH_dH_uH_d$ were to have little implication
for $CP$ and flavor-violating observables; the operator will of course provide corrections to the
sfermion and neutralino mass matrices, and can induce $CP$-odd mixing between $A$ and $h$, $H$, but
these effects do not lead to high sensitivity to $\La_h$. 

\begin{table}
\begin{center}
\begin{tabular}{c|c|c}
\hline
operator & sensitivity to $\La$ (GeV) & source \\ \hline\hline
$Y^{qe}_{3311}$ & $\sim 10^7$ & naturalness of $m_e$\\
Im($Y^{qq}_{3311})$ & $\sim 10^{17}$ & naturalness of $\bar \theta$, $d_n$\\
Im($Y^{qe}_{ii11}$) & $10^7-10^9$ & Tl, Hg EDMs \\ %\hline
$Y^{qe}_{1112}$, $Y^{qe}_{1121}$ & $10^7-10^8$& $\mu\rightarrow e$ conversion \\ %\hline
Im($Y^{qq}$) & $10^7-10^8$ & Hg EDM \\ %\hline
Im($y_h$) & $10^3-10^8$ & $d_e$ from Tl EDM \\ \hline
\end{tabular}
\end{center}
\caption{Sensitivity to the threshold scale. The naturalness bound on Im$(Y^{qq})$ 
doesn't apply to the axionic solution of the strong $CP$ problem, the best sensitivity to Im$(y_h)$ 
is achieved at maximal $\tan\beta$, and the Hg EDM constraint on Im$(Y^{qq})$ applies when at least one pair 
of quarks belongs to the 1$^{\rm st}$ generation.
}
\label{table1}
\end{table}

Remarkably enough, it turns out that EDMs do exhibit a high sensitivity to $H_uH_dH_uH_d$ 
at large $\tan\beta$ through corrections to the Higgs potential, and in particular the 
effective shift of the $m_{12}^2$ parameter,
%effective shift of the $m_{12}^2$ parameter in the Higgs potential 
%upon the inclusion of the correction (\ref{vh}) (we assume real $\mu$),
\be
m_{12}^2 H_u H_d \to (m_{12}^2)_{\rm eff}H_u H_d \equiv \left(m_{12}^2 + \fr{\mu y_hv_{SM}^2}{\Lambda_h}\right)H_u H_d.
\label{meff}
\ee
Crucially, a complex phase in $(m_{12}^2)_{\rm eff}$, due to Im$(y_h)$, is enhanced at large
$\tan\beta$ because $m_{12}^2 \simeq m_A^2/\tan\beta$. The resulting 
phase affects the one-loop SUSY EDM diagrams (see e.g. \cite{ourlateststuff}):
\ba
d_e&=&\fr{em_e\tan\beta }{16\pi^2m_{\rm susy}^2}\left (\fr{5 g_2^2}{24}+\fr{g_1^2}{24}\right)
\sin \left[{\rm Arg}\frac{\mu M_2}{(m_{12}^2)_{\rm eff}}\right].\;\;\;\;\;\;
\ea
Expanding to leading order in $1/\La_h$, using (\ref{meff}), and imposing the present limit on 
$d_e$ discussed earlier, one finds impressive sensitivity for large $\tan\beta$,
\be
\Lambda_h \ga 2\times 10^7 ~{\rm  GeV} \left(\fr{\tan\beta}{50}\right)^2
\left(\fr{300{\rm GeV}}{m_{\rm susy}}\right)\left(\fr{300{\rm GeV}}{m_A}\right)^2.
\ee

In conclusion, we have examined new flavor and $CP$ violating effects 
mediated by dimension five superpotential operators, and shown that the sensitivity to 
these operators extends far beyond the weak scale (as summarized in Table~1). The semileptonic operators that 
mediate flavor violation in the leptonic sector and/or break $CP$ could be 
detectable even if the scale of new physics is as high as $10^{9}$ GeV, and well above the naturalness 
scale. Our results can be translated into constraints on $CP$ and flavor violation 
in specific models leading to (\ref{qule}), {\em e.g.} the 
NMSSM or the MSSM with an extra pair of Higgses. Moreover, the sensitivity 
quoted in (\ref{cslimit}) and (\ref{muelimit}) is robust, having a mild dependence 
on the SUSY threshold.  
Finally, since these effects decouple linearly, an increase in sensitivity by 
just two orders of magnitude would already start probing scales relevant for neutrino physics.  
Our results motivate further searches for EDMs and LFV in the SUSY framework 
even if the soft-breaking sector provides no new sources, as happens {\em e.g.} in models with low scale SUSY breaking.

\end{document}